# Trapping and unlimited delay of light pulses at microscale without distortion

M. Sumetsky

Aston Institute of Photonic Technologies, Aston University, Birmingham B4 7ET, UK

m.sumetsky@aston.ac.uk

**Abstract**

A tunable bottle microresonator can trap an optical pulse of the given spectral width, hold it as long as the material losses permit, and release without distortion.

1. **Introduction**

A microscale optical buffer (MOB) is proposed as one of the key components of the future optical signal processors on the chip [1, 2]. MOB traps the input pulse, holds it over the predetermined period of time, and eventually releases without distortion. Combination of the smallest possible dimensions of the buffer with the required delay times suggests that light residing in an MOB experiences multiple reflections or rotations before it is finally released. This causes distortion of the pulse due to dispersion, multiple self-interferences, and attenuation. While the direct coiling of photonic waveguides allows one to come up with compact and broadband delay lines [3, 4] much smaller MOBs can be fabricated based on optical microresonators and photonic crystals. The latter structures though suffer from the bandwidth-delay time limitation [1, 2]. A way to overcome this limitation suggested in [5] consists in creation of a MOB by adiabatic compression of the transmission band of a coupled resonator optical waveguide (CROW). Ideally this CROW MOB enables slowing down and stopping of a light pulse with the predetermined spectral width. However, significant practical barriers, such as insufficient precision of modern photonics technologies and attenuation of light, remain in the way of fabrication of such MOBs. In this paper, an alternative feasible MOB is proposed and demonstrated by numerical calculations. In contrast to the previously suggested MOBs based on microresonators and photonic crystals [1-3], the device considered here is a specially designed tunable bottle resonator illustrated in Fig. 1. The effective radius variation of this resonator is dramatically small and has a parabolic shape of a few nanometer height only. This variation can be introduced along an optical fiber with the unprecedented sub-angstrom precision using the Surface Nanoscale Axial Photonics (SNAP) platform [6, 7]. The device proposed here is a tunable generalization of the record small and low loss bottle resonator delay line experimentally demonstrated in [7]. The shape of an optical pulse bouncing along the axis of such resonator with parabolic radius variation changes periodically and, thus, is fully recovered after a roundtrip without distortion. Based on this result, which was found by Schrödinger in the early days of quantum mechanics [8], it is demonstrated that the appropriate nanoscale temporal and spatial tuning of the effective radius of this resonator allows to trap, hold, and release a telecommunication pulse without distortion for the period of time limited by the material losses only.

## 2. Trapping, holding, and release of an optical pulse

The bottle resonator considered below is an optical fiber segment with nanoscale effective radius variation illustrated in Fig. 1(a). Light is coupled in and out of this resonator through a transverse waveguide (microfiber taper). The buffering process includes: opening the bottle resonator by nanoscale variation of its effective radius (refractive index) and entering the optical pulse (Fig. 1(b)); closing the resonator when the pulse is completely inside it and holding the pulse in the parabolic bottle resonator for the duration of the required time delay (Fig. 1(c)); and releasing the pulse by reversing the deformation illustrated in Fig. 1(b) (Fig. 1(d)).

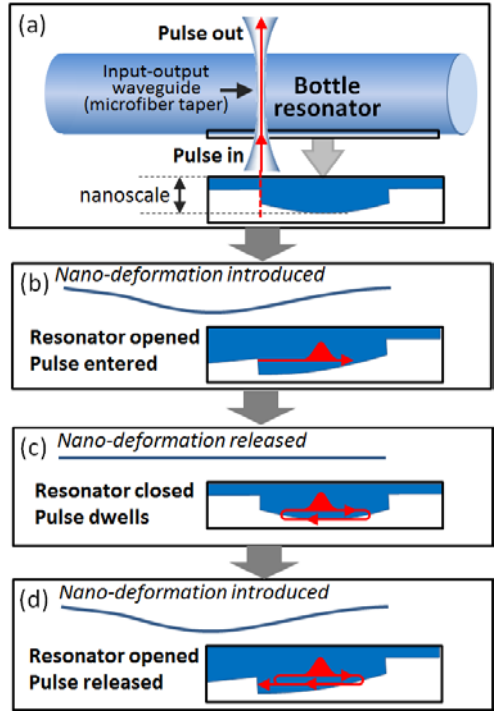

**Fig. 1.** (a) – Illustration of the nanoscale radius variation of a parabolic bottle resonator, which is magnified in the inset. (b) – The resonator is opened by the nanoscale variation of the effective fiber radius to let the input pulse in. (c) – After this variation is released, the captured optical pulse bounces inside the parabolic resonator. (d) – The pulse is coupled out by application of the same nanoscale deformation.

Due to the very small and smooth effective radius variation inside the bottle resonator considered, the propagation of a whispering gallery mode (WGM) can be fully described by its amplitude variation $\Psi(z,t)$ along the fiber axis $z$. The equation that describes this propagation has the form of the one-dimensional Schrödinger equation [6, 7]. For the non-stationary case under consideration this equation takes the form:

$$i\mu\frac{\partial \Psi}{\partial t} = -\frac{\partial^2 \Psi}{\partial z^2} + V(z,t)\Psi . \qquad (1)$$

Here $\mu = 2kn/c$ and $V(z,t) = -2k^2\Delta r_{eff}(z,t)/r_0$ are defined through the radiation wavelength $\lambda$ of the transmission channel, refractive index $n$ and bulk propagation constant $k = 2\pi n/\lambda$ of the bottle resonator material, speed of light in vacuum $c$, and the nanoscale effective variation $\Delta r_{eff}(z,t)$ of the fiber radius $r_0$.

We first consider the propagation of a telecommunication pulse launched in the middle of a rectangular bottle resonator having the height $\Delta r_0 = 4$ nm and length 2 mm (Fig. 2(a)). Here and below, we consider the propagation of a 100 ps pulse and set the fiber radius $r_0 = 20$μm, refractive index $n = 1.5$, and wavelength $\lambda = 1.5$ μm. The surface plot describing the evolution of this pulse along the fiber axis $z$ as a function of time is shown in Fig. 2(b). It is seen that the pulse experiences significant corruption in the process of bouncing caused by both dispersion and self-interference [9]. As the result, the original shape of the pulse is completely lost in a few nanoseconds.

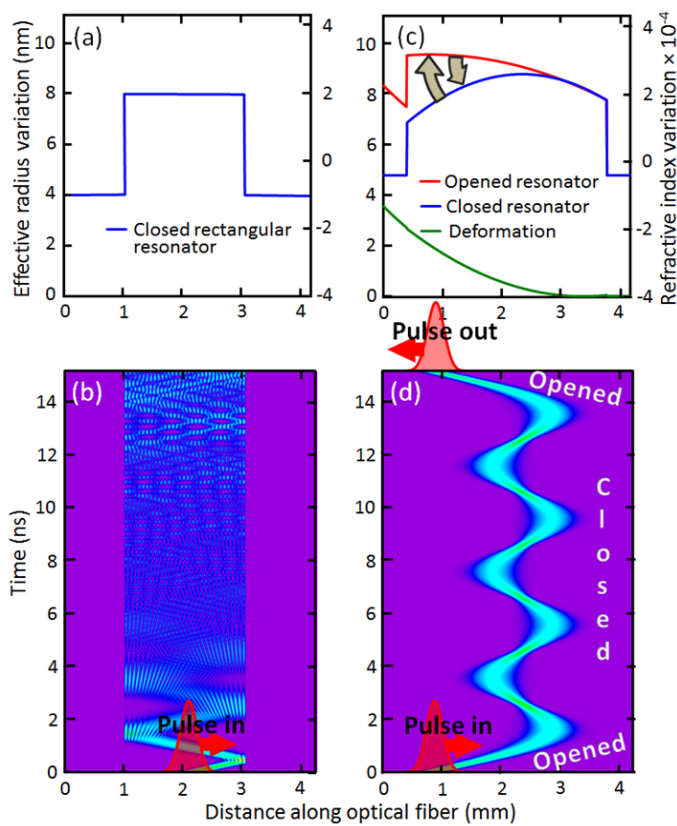

Fig. 2. Pulse propagation in bottle resonators having the rectangular, (a), (b), and parabolic, (c), (d), shapes. Figs. 2(a) and (c) show the profile of the effective radius variation (left axis) rescaled to the refractive index variation (right axis) of these bottle resonators. Respectively, Figs. 2(b) and (d) show the evolution of the 100 ps light pulse in these resonators.

On the contrary, multi-period bouncing of an optical pulse in a bottle resonator with a parabolic effective radius variation does not cause irreversible deformation of the pulse. This fact follows from the periodicity of motion of a wave packet described by the Schrödinger equation (1) with a parabolic potential [8]. The process warranting full restoration of the input pulse after multiple bouncing in the bottle resonator is illustrated in Figs. 2(c) and (d). The original bottle resonator with 2 nm height (blue curve) is opened by the introduced deformation ~ 3 nm (or refractive index variation ~$3 \cdot 10^{-4}$) to enter the 100 ps pulse shown on the bottom of Fig. 2(d). The deformed resonator has a semi-parabolic shape, which allows us to avoid the pulse dispersion during the entering, exiting, and switching periods [7]. The deformation is gradually released without perturbing the pulse during the time period 1 ns to 2 ns when the pulse is situated near the right hand side of the resonator. After the switching process is finished, the pulse is forced to bounce inside the resonator as long as the resonator is closed (the middle part of Fig. 2(d)). Finally, the resonator is gradually opened during the time between 13 ns and 15 ns to completely release the pulse. Comparison of the shapes of the input and output pulses shows no distortion.

## 3. Summary and discussion

A few nanometer tuning of the bottle resonator effective radius (equivalent to the variation of its refractive index $\sim 10^{-4}$) is sufficient to trap, hold, and release telecommunication optical pulses without distortion over the time period of tens of nanoseconds or longer, while the delay time is limited by the material losses only. The dimensions of the MOB are determined by the footprint of the bottle resonator (0.08 mm$^2$ for the model considered; each oscillation of pulse in this resonator delays light by 3.5 ns). The switching function of the proposed MOB can be realized by tuning the nanoscale deformation (refractive index) of the bottle resonator with a piezoelectric transducer attached to the fiber or coating it in the vicinity of the bottle resonator. This deformation can also be introduced by a nanosecond laser pulse which intensity is appropriately distributed along the fiber length. The power of the switching pulse can be enhanced dramatically if this pulse is resonantly coupled into the fiber WGM [10]. Since the required spatial distribution of the switching deformation is smooth (Fig. 2 (c)), the corresponding intensity variation is feasible. Alternatively, the required nanoscale temporal and spatial variation can be introduced by in the fiber fabricated of a low loss piezoelectric or electrostrictive material, polled silica fiber, the silica fiber with a piezoelectric or electrostrictive core [11], and external acousto-optic or electro-optics by application of laser or electric pulses. The recently demonstrated WGM all-optical tuning in a silica fiber having a silicon core [12] confirms the feasibility of the proposed microscopic optical buffer. Remarkably, the parabolic profile of the bottle resonator is not the unique profile which allows to hold the light pulse without noticeable distortion. In fact, there exists a wide family of potential wells in which the period of classical oscillations does not depend on its amplitude [13]. Using the bottle profiles determined by these generalized potentials allows us to make the design of MOBs more flexible. In addition, the advanced manipulation of optical pulses by complex deformation of the bottle resonator in the process of delay is of special interest.